\documentclass[conference]{IEEEtran}
\IEEEoverridecommandlockouts
\usepackage{cite}
\usepackage{amsmath,amssymb,amsfonts}
\usepackage{algorithmic}
\usepackage{graphicx}
\usepackage{svg}
\usepackage{textcomp}
\usepackage{pgfplots}
\pgfplotsset{compat=1.16} 
\usepackage{colortbl}
\usepackage{url}
\usepackage{siunitx}
\usepackage{tabularx}
\usepackage[hidelinks]{hyperref}
\usepackage{booktabs}
\usepackage[T1]{fontenc} 
\usepackage[normalem]{ulem}
\def\BibTeX{{\rm B\kern-.05em{\sc i\kern-.025em b}\kern-.08em
    T\kern-.1667em\lower.7ex\hbox{E}\kern-.125emX}}

\usepackage{listings}

\colorlet{punct}{red!60!black}
\definecolor{background}{HTML}{EEEEEE}
\definecolor{delim}{RGB}{20,105,176}
\colorlet{numb}{magenta!60!black}

\lstdefinelanguage{json}{
    basicstyle=\normalfont\ttfamily,
    numberstyle=\scriptsize,
    stepnumber=1,
    numbersep=8pt,
    showstringspaces=false,
    breaklines=true,
    frame=lines,
    backgroundcolor=\color{background},
    literate=
     *{0}{{{\color{numb}0}}}{1}
      {1}{{{\color{numb}1}}}{1}
      {2}{{{\color{numb}2}}}{1}
      {3}{{{\color{numb}3}}}{1}
      {4}{{{\color{numb}4}}}{1}
      {5}{{{\color{numb}5}}}{1}
      {6}{{{\color{numb}6}}}{1}
      {7}{{{\color{numb}7}}}{1}
      {8}{{{\color{numb}8}}}{1}
      {9}{{{\color{numb}9}}}{1}
      {:}{{{\color{punct}{:}}}}{1}
      {,}{{{\color{punct}{,}}}}{1}
      {\{}{{{\color{delim}{\{}}}}{1}
      {\}}{{{\color{delim}{\}}}}}{1}
      {[}{{{\color{delim}{[}}}}{1}
      {]}{{{\color{delim}{]}}}}{1},
}

\definecolor{success}{HTML}{98B4D4}
\definecolor{fail}{HTML}{E15D44}

\begin{document}

% \title{A Contract-Based Intent Framework for Agentic Slicing in OpenRAN\\
\title{Contract-based Agentic Intent Framework for Network Slicing in O-RAN\\
}

% \author{\IEEEauthorblockN{
%         Fransiscus Asisi Bimo\IEEEauthorrefmark{1},
%         Maria Amparo Canaveras Galdon\IEEEauthorrefmark{2}\IEEEauthorrefmark{3},
%         Chun-Kai Lai\IEEEauthorrefmark{1},\\
%         Ray-Guang Cheng\IEEEauthorrefmark{1},
%         and
%         Edwin K. P. Chong\IEEEauthorrefmark{3}
%        \\
%   }
% \IEEEauthorblockA{\IEEEauthorrefmark{1}
% National Taiwan University of Science and Technology, Taiwan
%   }
% \IEEEauthorblockA{\IEEEauthorrefmark{2}
% NVIDIA, USA
%   }
% \IEEEauthorblockA{\IEEEauthorrefmark{3}
% Colorado State University, USA  \\
%   }
%   Email: crg@mail.ntust.edu.tw
% }

\author{\IEEEauthorblockN{
        Fransiscus Asisi Bimo\IEEEauthorrefmark{1},
        Chun-Kai Lai\IEEEauthorrefmark{1},
        Zhi-Yuan Yang\IEEEauthorrefmark{1},
and
        Ray-Guang Cheng\IEEEauthorrefmark{1}
       \\
  }
    \IEEEauthorblockA{\IEEEauthorrefmark{1}
    Dept. of Electronic and Computer Engineering, National Taiwan University of Science and Technology, Taiwan
    }  
    
    Email: crg@mail.ntust.edu.tw,
    }

\maketitle

\begin{abstract}
Intent-based networking aims to simplify network operation by translating operator intents into a collection of policies, configurations, and control actions. However, this translation process relies on heuristics and loose coupling. It often results in unpredictable behavior and ambiguous safety standards.
This paper presents a Contract-based Agentic Intent Framework (CAIF) for the radio access network (RAN). The proposed framework employs a closed-loop agentic pipeline that systematically audits user objectives against formal RAN constraints prior to actuation. The proposed CAIF decouples probabilistic intent extraction from strictly governed policy execution to enable the enforcement of deterministic safety guarantees. We use network slicing as a representative use case to demonstrate the design flow and validate the effectiveness of the proposed approach on an O-RAN testbed. Experimental results show that the closed-loop agentic
pipeline of the proposed CAIF can effectively eliminate harmful intent executions observed in direct-actuation baseline approaches.

\end{abstract}


\section{Introduction}

The transition to 6G networks introduces an era of network flexibility, where the management of heterogeneous resources like Radio Resource Management (RRM) requires a fundamental shift toward full autonomy.  Intent-Based Networking (IBN) has emerged as the essential paradigm, allowing operators to define high-level operational goals without managing low-level parameters. This autonomy is particularly critical for Network Slicing \cite{9443201}, a complex mechanism requiring the dynamic, cross-domain partitioning of resources to meet strict Service Level Agreements (SLAs). However, realizing the full potential of IBN for slicing remains challenging due to the rigid nature of traditional translation models. In this context, Large Language Models (LLMs) present a transformative opportunity, offering a cognitive layer capable of interpreting and translating high-level human intent and decomposing it into actionable logic\cite{multiagentreference}\cite{10574890}.

Translating a high-level goal into a specific outcome requires precise, granular control over network resources. It requires a programmable and modular infrastructure where every component can be individually managed via software. Programmable RAN architectures\cite{9814869} serve as the necessary execution substrate by decoupling control logic from physical radio hardware through open and standardized interfaces. This separation transforms the network from a monolithic black box into a modular platform where control functions are exposed as software applications, specifically Non-Real-Time (rApps) and Near-Real-Time (xApps). By utilizing this disaggregated framework, high-level slicing requirements can be programmatically mapped to specific control loops, providing the granular, evidence-based actuation layer necessary to partition radio resources dynamically in a real-world environment.

A prior study implementing an intent-driven mechanism for O‑RAN automation demonstrates strong potential in orchestrating multiple xApps to optimize key performance indicators (KPIs) such as throughput and energy efficiency\cite{10298342}. However, it still rely on manually specified operator intents. The system requires operators to input explicit KPI targets. This approach lacks an intent‑interpretation layer capable of translating human‑level semantic instructions into machine‑readable optimization goals. 
To address these constraints, we propose the integration of LLMs as a flexible intent-interpretation layer. However, we anticipate that this may introduce architectural risks that require careful mitigation.

Deploying LLM-based agents introduces inherent reliability risks due to their probabilistic nature, where outputs may drift even under static inputs \cite{11152698}. While methods like JSON schemas enforce formatting \cite{10539172}, they fail to validate decision logic, leaving the network unable to distinguish between valid optimizations and dangerous fabrications. Without strict limits, these agents can react too aggressively to a request, causing a chain reaction of errors that crashes the wider network\cite{10327837}. Our prior work confirmed this vulnerability, showing that agents often yield inconsistent interpretations despite accurate parsing \cite{bimo2025intentbasednetworkranmanagement}. To evaluate this problem, we adopt this earlier architecture as our Direct-Actuation Baseline, using it to demonstrate that without a formal Intent Contract as a deterministic guardrail, autonomous decisions remain unsafe to execute.

To address these reliability issues, we present a Contract-based Agentic Intent Framework specifically built for slicing in O-RAN networks. Our contributions are as follows:

\begin{itemize} \item Contract-based Agentic Intent Framework Architecture: We design a translation pipeline where the LLM output is strictly governed by a formal Intent Contract which enforces deterministic validation before execution. \item Slicing Specific App Orchestration: We demonstrate how high-level intent contracts are systematically decomposed and enforced across a hierarchical rApp and xApp workflow in a programmable RAN, enabling dynamic and SLA aware network slicing. \item Implementation in Open RAN Architecture: We validate the practical feasibility of the proposed framework on an Open RAN testbed, showing how natural language intents are translated into standardized rApp or xApp control policies under contract enforcement. \end{itemize}

The manuscript is organized as follows. Section \ref{Proposed Architecture} discusses the proposed architecture. Section \ref{Experiment Result} discusses the results of the experiment. Finally,
Section \ref{Lesson Learned} addresses the lesson learned from our experiment and potential future works.

\begin{table*}[htbp] % Use table* to span both columns
    \caption{Mandatory KPIs and Constraints Defined in Intent Contract}
    \label{tab:intent_kpis}
    \centering
    \begin{tabular}{l l l} % You might want to use {l l p{5cm}} if text wraps
        \hline
        \textbf{KPI / Parameter} & \textbf{Value / Constraint} & \textbf{Source Context} \\
        \hline
        Target Resource & \texttt{Cell\_1\_Slice\_1} & \textit{icm:target} \\
        Expectations & Throughput Enhancement & \textit{icm:hasExpectation} \\
        Target Value & 5\% & \textit{ran:targetThroughputIncreasement} \\
        Policy Mechanism & Two-level RRM Policy Ratio & \textit{idan:Delivery\_slaPolicy} \\
        IntentSpecification & slaSliceSpec & \textit{id} \\
              IntentRelationship & policy-baseline & \textit{id} \\
        characteristic & [eligibleClusters, affectedCells, string, Cell\_1\_Slice\_1] & \textit{[id, name, valueType, value]} \\
        \hline
    \end{tabular}
\end{table*}

\section{Contract-based Agentic Intent Framework} \label{Proposed Architecture}
This paper proposes a Contract-based Agentic Intent Framework (CAIF) designed to operate within the disaggregated O-RAN architecture. Our method uses Generative AI to understand complex requests. However, to prevent errors, we require the AI to assist in generating a \textit{digital contract} that verifies the safety of its decision before the network executes it. The architecture establishes a modular pipeline where high-level intent is first processed into a formal contract before being executed via standardized O-RAN interfaces.

\subsection{Contract-based Intent Model} Our framework integrates the TMF-921 Intent Management API \cite{tmf} to create a formal \textit{Intent Contract}. To ensure deterministic execution, we enforce a strict schema where abstract goals are mapped to specific JSON-LD paths. As shown in Table \ref{tab:intent_kpis}, the contract mandates specific KPIs that strictly define the agent's operating boundary prior to actuation.

The contract consists of mandatory KPIs that the agentic framework must populate and validate:
\begin{itemize}

    \item Scope and Expectations: The icm:target and icm:hasExpectation fields explicitly define the scope of the operation. In the example shown, the intent is strictly limited to Throughput Enhancement on Cell\_1\_Slice\_1, preventing the agent from inadvertently modifying adjacent resources or applying incorrect optimization logic.

    \item Deterministic Constraints: Unlike open-ended LLM outputs, the contract enforces quantifiable limits. The Target Value (e.g., 5\%) sets a rigid boundary for the optimization, ensuring that the ran:targetThroughputIncreasement does not exceed safe operational thresholds.

    \item Policy Mechanism: The idan:Delivery\_slaPolicy field specifies the exact control loop to be utilized—in this case, the Two-level RRM Policy Ratio. This decoupling ensures that while the LLM determines what needs to happen, the underlying standardized rApps determine how it is executed, preventing the generation of hallucinated or non-standard command sequences.

    \item Governance and Relationships: To ensure lifecycle integrity, the contract links to a specific IntentSpecification (slaSliceSpec) and establishes an IntentRelationship with the policy-baseline. This hierarchy allows the Intent Management system to detect conflicts with existing policies before they are activated.

    \item Flattened Characteristics: For rapid indexing and retrieval by rApps/xApps, complex graph attributes are flattened into the characteristic array (e.g., eligibleClusters), allowing low-latency components to access vital parameters without parsing the full semantic graph.

\end{itemize}
\subsection{Agentic AI for Dynamic Translation} 
CAIF adopts a dual-agent architecture to support dynamic and reliable intent translation.
The overall system consists of four core components, which are detailed as follows:

\begin{itemize}
    \item Profiling Agent: Analyzes multi-turn natural language inputs and extracts structured intent information according to a predefined intent model. It determines the intent category and key parameters while ensuring schema compliance.
    \item Evaluator Agent: Validates the extracted intent by checking completeness, consistency, and constraint compliance. If violations or missing fields are detected, it triggers refinement to improve translation reliability.
    \item Contract Generator:  Transforms the validated intent into a standardized, machine-readable contract. The generated contract is suitable for execution, validation, and auditing.
    \item Intent Management:  Registers and tracks contracts through their entire lifecycle. It also maintains the \textit{IntentSpecifications} that define what the system is allowed to do.
\end{itemize}

\subsection{Prompt Techniques} \label{Prompt}

\begin{figure}
    \centering
    \includegraphics[width=1\linewidth]{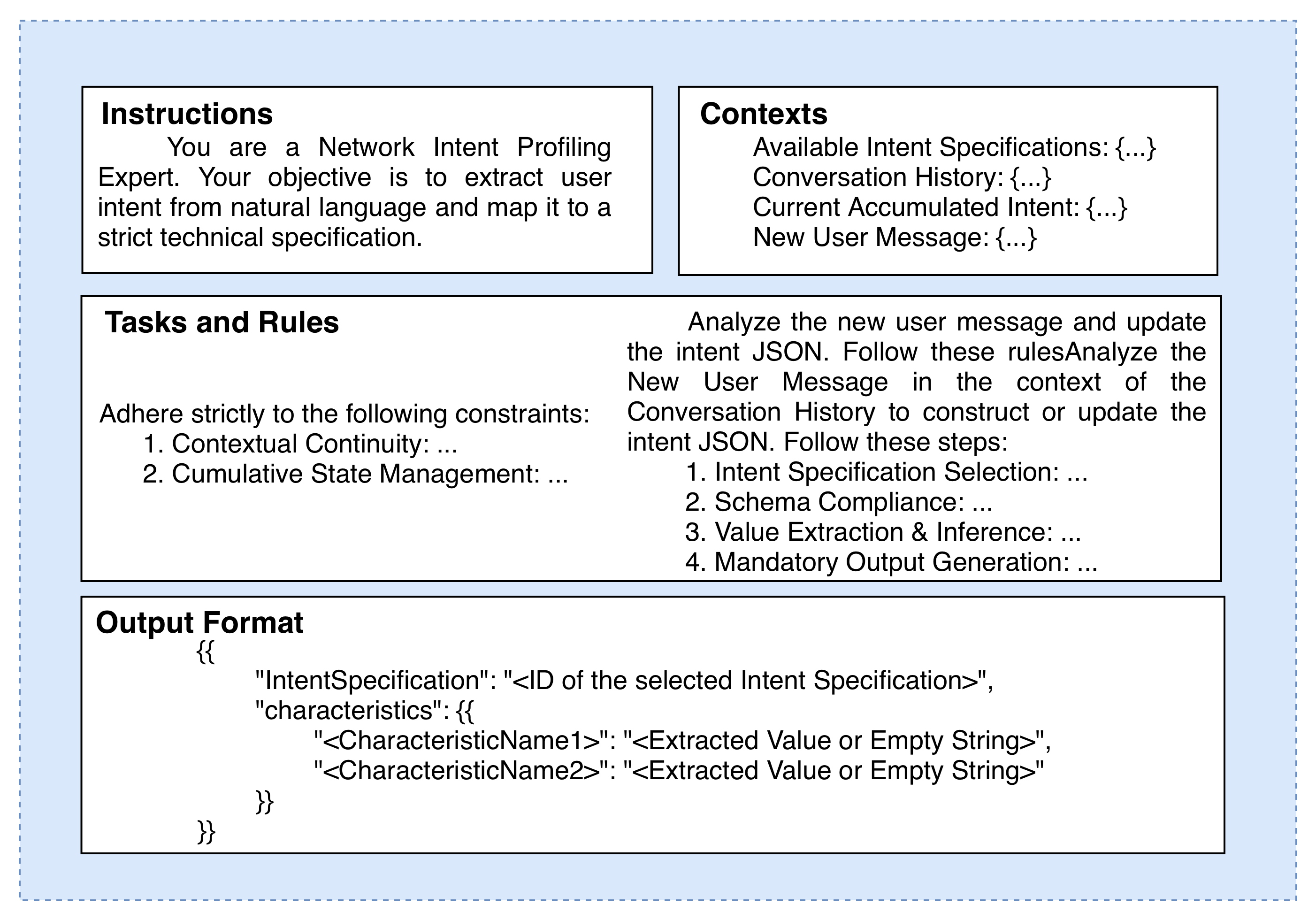}
    \caption{Intent Profiling Prompt}
    \label{fig:profilingprompt}
\end{figure}

We design three types of prompts, namely the Intent Profiling Prompt, Intent Evaluation Prompt, and Intent Refinement Prompt, to achieve reliable and controllable intent extraction using LLMs. The three prompts share a similar structure as shown in Fig. \ref{fig:profilingprompt}. The structure constrains the LLM to the telecommunications domain and applies iterative filling together with strict schema enforcement to translate multi-turn natural language inputs into machine-readable intents. To handle the unpredictability of LLM outputs, the Intent Evaluation Prompt acts as a strict content validator. This mechanism cross-checks the extracted intent against the full conversation history to ensure the accuracy of intent. Following this, the Intent Refinement Prompt triggers a correction loop. Instead of rewriting the entire request, this step fixes only the specific errors identified by the evaluation and preserves the parts of the intent that were already correct.

\subsection{Intent Translation Pipeline}
Utilizing the prompt strategies from Sect. \ref{Prompt}, we construct an agentic pipeline to translate semantic intents into contracts. This architecture ensures semantic accuracy through agentic reasoning and iterative self-correction. The system relies on two specialized agents: the Profiling Agent, which handles intent parsing and structuring, and the Evaluator Agent, which performs error detection and semantic review.

The workflow starts with the Profiling Agent, which translates the operator's input into a structured format derived from the intent specifications. The Evaluator Agent then inspects this output to detect inconsistencies, hallucinations, or unsupported requests. If anomalies are identified, the system triggers a closed-loop refinement process, guiding the Profiling Agent to rectify the specific discrepancies based on feedback. Once the output is fully validated, the Contract Generator produces the final Intent Contract. This systematic verification ensures that strictly deterministic instructions are enforced prior to network execution.

% Our system is built upon two distinct agents: the History Analyzer Agent and the Strategist Agent. The History Analyzer Agent is responsible for analyzing past strategies applied in the network and obtains real-time measurements directly from the RAN as input. Data about past strategy attempts are obtained from the Historical Strategy Attempts database. The History Analyzer Agent recommends the best strategy outcomes to the Strategist Agent. This processed data, along with a number of the most relevant previous strategy attempts, is then passed to the Strategist Agent for decision-making. Predefined strategy that contains data about target cell capabilities stored in the Strategy Templates database. Both agents perform their respective inference tasks calling the NVIDIA NIM inference API with detail inference as summarized in Table \ref{tab:hyperparameters}. The underlying intelligence for these agents is configured within the Prompt Processor component, which utilizes the LangGraph library \footnote{https://github.com/langchain-ai/langgraph}.

\subsection{CAIF Implementation on the O-RAN Platform} 

\begin{figure*}
    \centering
    \includegraphics[width=0.9\linewidth]{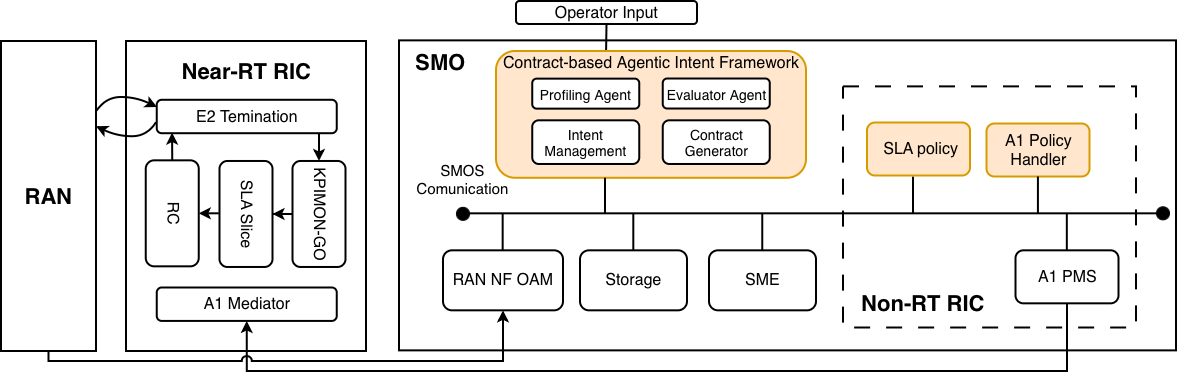}
    \caption{System architecture of a CAIF}
    \label{fig:intent-system}
\end{figure*}

We implement the CAIF for network slicing within an O-RAN environment, as illustrated in Fig. \ref{fig:intent-system}. To ensure a realistic testing environment, our system utilizes an experimental platform similar to the one described in \cite{10464635}. This setup allows us to demonstrate the practical execution of Intent Contracts through specific rApp and xApp service calls.

\subsubsection{SLA Slice and A1 Policy Handler}
Once the SLA Slice rApp receives the formal intent, it converts the user request from a percentage into a absolute throughput threshold. For example, if a user asks for a percentage change, the system calculates exactly how many megabits per second that represents. This calculation uses both live and past network data collected from the O-RAN O1 interface within the SMO.

After setting this target, the SLA Slice rApp performs a feasibility check to see if the network can actually reach that goal. It does this by looking at a table of past performance that tracks Physical Resource Block (PRB) and Channel Quality Indicator (CQI) of a cell. By comparing the new request to this history, the system decides if the target is possible under the current radio conditions. Only requests that pass this check move forward to the next step in the control process.
% \textcolor{blue}{After receiving a formalized Intent, the SLA Slice rApp first translates the Intent into a quantifiable performance target which from percentage to an absolute value. This conversion is based on real-time and historical network performance data collected through the O-RAN O1 Performance Management (PM) interface via RAN NF Operations and Maintenance (OAM) function components in SMO.}

% \textcolor{blue}{The SLA Slice rApp then performs a feasibility check to see if the network can actually reach the requested goal. To do this, it looks at a pre established table that shows past performance which records PRB and CQI of all cells. By comparing the user request to this historical data, the system decides if the target is possible under current conditions. Only requests that pass this check move forward to the next step in the control process.}

After a successful check, the SLA Slice rApp creates the necessary A1 policy data and sends a unique policy name back to the CAIF. Once the A1 Policy Handler rApp receives the notification, it opens and reads the full Intent Contract to find the matching policy data. It then sends this information to the A1 Mediator on the Near-RT RIC using the A1 Policy Management Service within the SMO. 

\subsubsection{Closed-Loop RAN Slice Control with E2 KPM and RC}
To execute the policies generated by the CAIF, we implement a closed-loop control system within the Near-RT RIC. This workflow relies on the interaction among three specialized xApps to maintain the required slice performance. First, the KPIMON-GO xApp continuously monitors the network by collecting key performance measurements and storing them in a database.

\begin{figure}
    \centering
    \includegraphics[width=1\linewidth]{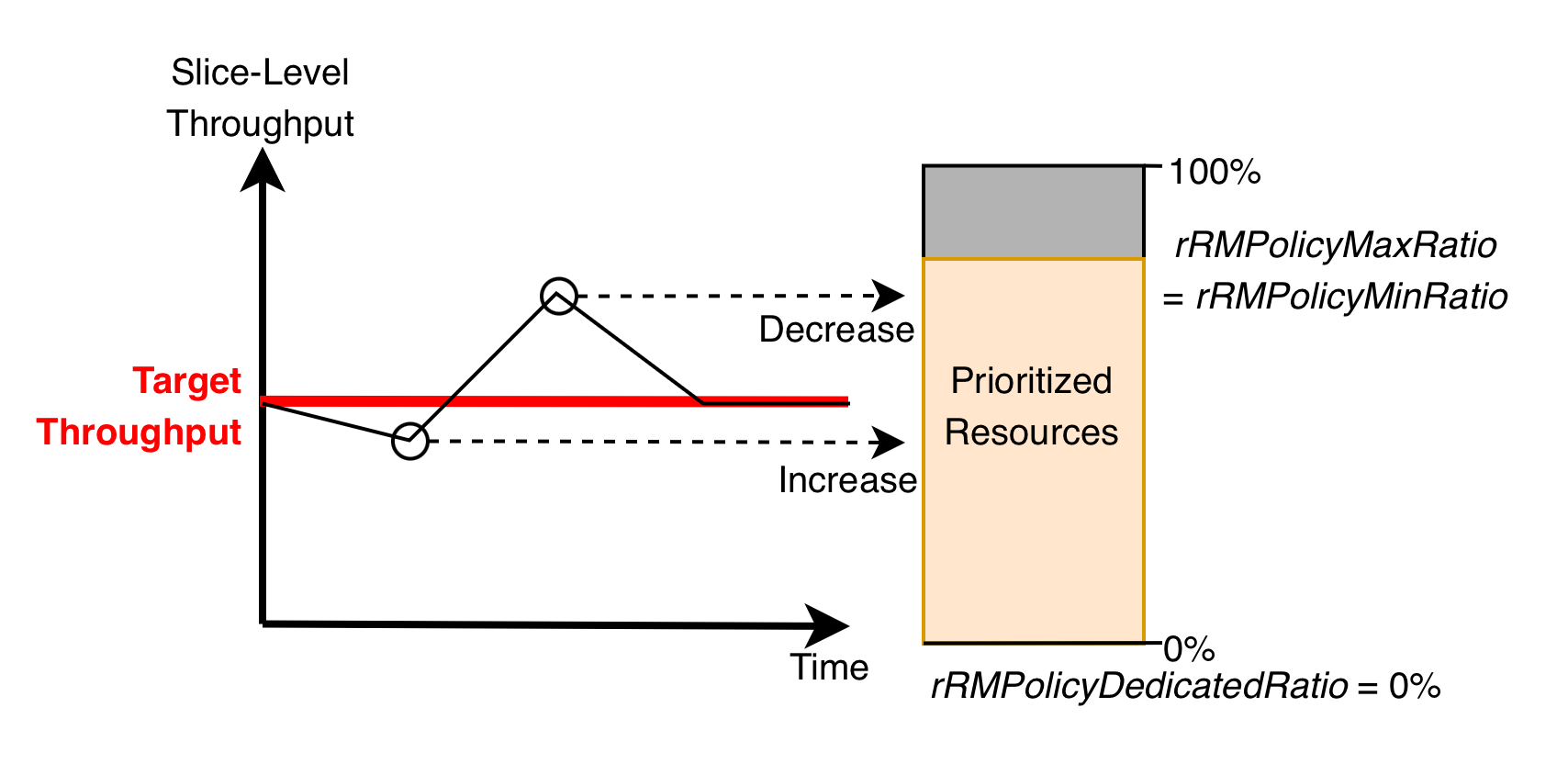}
    \caption{Closed-Loop RAN Slice Control Algorithm}
    \label{fig:slaslicexapp}
\end{figure}

The SLA Slice xApp then retrieves this data to evaluate slice performance against the defined contract. If the current throughput does not meet the goal, this app dynamically calculates new values for the maximum and minimum resource ratios as shown in Fig. \ref{fig:slaslicexapp}. These decisions are passed to the RC xApp, which functions as the execution unit. The RC xApp converts these high-level control decisions into encoded messages that are sent to the RAN.

% \subsection{Closed-Loop RAN Slice Control with E2 KPM and RC}
% \textcolor{blue}{We implement a closed-loop RAN slice control to fulfill the policies provided by the Intent-Based Network. It is implemented in the Near-RT RIC, which includes the Near-RT RIC Platform and multiple xApps connected to RAN nodes through the E2 interface. The E2 interface provides standardized service models for RAN management. This work uses the RAN Control (RC) service model and the Key Performance Measurement (KPM) service model~\cite{oran_e2sm_201} to enable control actions and KPI collection }

% \textcolor{blue}{Based on these E2 service models, we implement the KPIMON-GO xApp, the SLA Slice xApp, and the RC xApp to support the closed-loop RAN slice control. The KPIMON-GO xApp is a Key Performance Indicator(KPI) monitoring xApp responsible for collecting key performance measurements and storing them in a database via the E2 interface using the KPM service model~\cite{oran_kpm_203}. The SLA Slice xApp monitors the performance data retrieved from the database and dynamically adjusts the \textit{rRMPolicyMaxRatio} and \textit{rRMPolicyMinRatio} of a slice within a specific cell to allocate the prioritized resource (Fig. \ref{fig:slaslicexapp}), ensuring that the target throughput complies with the policies provided by Non-RT RIC through the A1 interface. The RC xApp operates as a RAN control xApp that parses control decisions from the SLA Slice xApp into the ASN.1-encoded E2 messages defined in~\cite{oran_rc_103}.}

\section{Experimental Results}  \label{Experiment Result}
\subsection{Experimental Setup and Methodology}
The evaluation of the proposed contract-based framework is conducted using the O-RAN experimental platform and testbed infrastructure established in our previous work \cite{10464635}. This platform provides the necessary cell configuration and measurement functions to simulate a disaggregated RAN environment. For RAN itself, we utilize the commercial AI RSG RAN simulator from VIAVI. For this study, we focus on the reliability of the intent to policy translation process. We generate a curated and custom dataset of 500 natural language intent instances, with variations across different shot cases ranging from 1 to 5, which is publicly available at \cite{intent_dataset}. This variation covers a wide range of possibilities for user intent. These instances feature diverse semantic structures and parameter values, such as varying slice identifiers and throughput thresholds.

\subsection{Semantic Integrity and Output Validation} 
\begin{figure}
    \centering
    \includegraphics[width=1\linewidth]{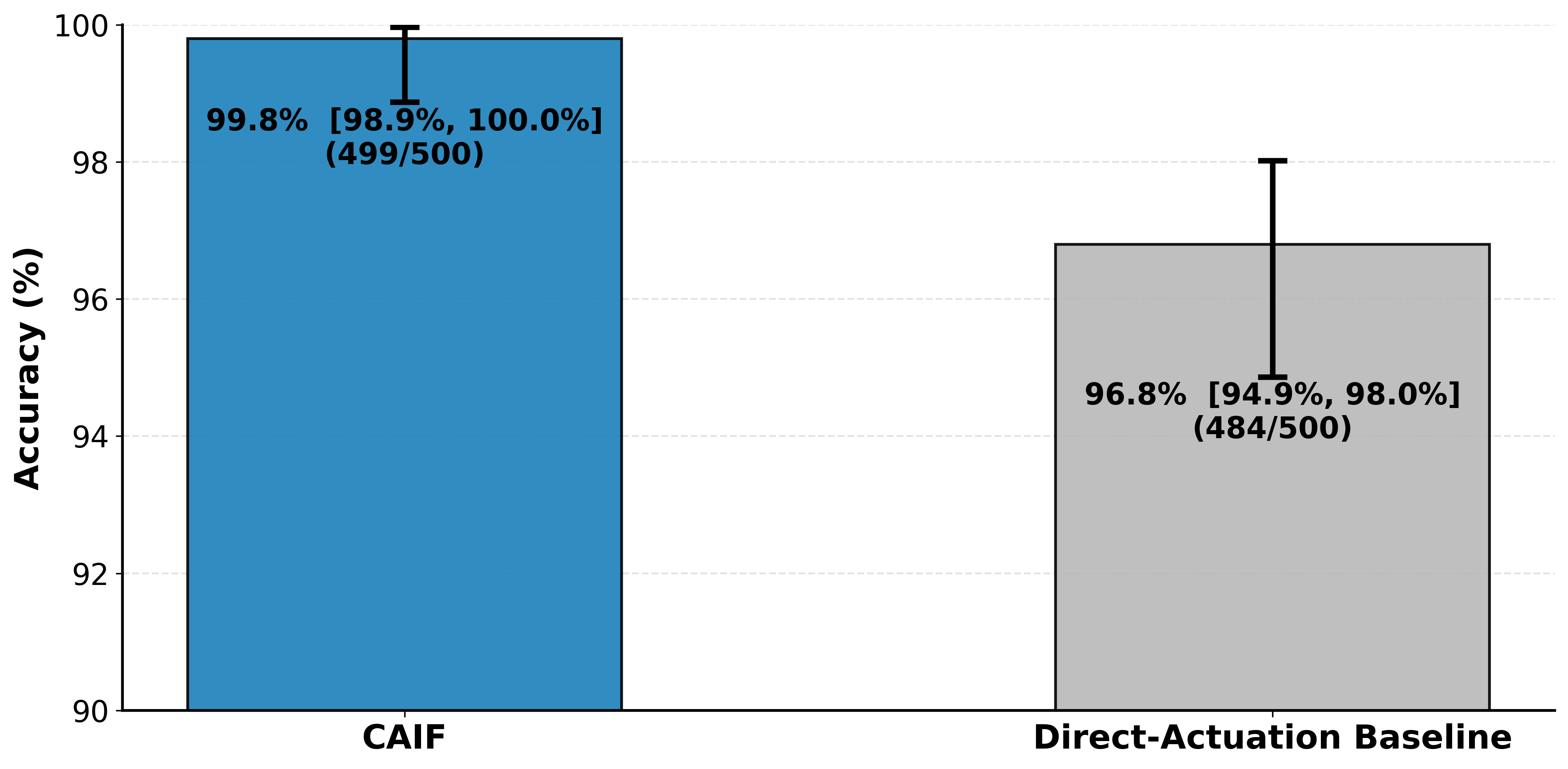}
    \caption{Overall Accuracy with 95\% Confidence Intervals}
    \label{fig:SSSO_vs_aLLM}
\end{figure}

This evaluation assesses the semantic integrity of the intent-to-policy translation process. We observed whether the output generated by the baseline LLM or CAIF can align with the original user intents, or if semantic deviation occurred during translation. As illustrated in Fig. \ref{fig:SSSO_vs_aLLM}, the baseline architecture (configured as Table \ref{tab:hyperparameters_baseline}) achieves a translation accuracy of 96.8\%. However, its 95\% confidence interval ranges from 94.9\% to 98.0\%, indicating a higher risk of semantic inconsistency. From the 500 test cases, 16 errors were observed in the baseline results. Some of the errors identified as false positives, where the generated A1 policies were structurally valid yet semantically misaligned with the user’s intent. This error category is particularly dangerous because these policies pass basic syntax checks and are therefore forwarded for execution, directly violating the operational requirements. The remaining failures were caused by formatting errors, where the output did not match the strict requirements of the rApp. In the baseline architecture, these malformed commands are sent directly to the network, leading to guaranteed execution errors.

\begin{table}[h!]
    \centering
    \caption{Direct-Actuation Baseline LLM Agent Inference Parameters}
    \label{tab:hyperparameters_baseline}
    \begin{tabular}{ cc }
        \toprule % This will be a bold/thick line at the top
        \textbf{Parameter} & \textbf{Value} \\
        \midrule % This will be a standard mid-rule
        LLM Agent Model & llama-3.3-nemotron-super-49b-v1.5 \\
        Temperature & 0.6 \\
        Top P & 0.95 \\
        \bottomrule % This will be a bold/thick line at the bottom
    \end{tabular}
\end{table}

In contrast, CAIF (configured as Table \ref{tab:hyperparameters_CAIF}) achieves an accuracy of 99.8\%, with a much narrower 95\% confidence interval ranging from 98.9\% to 100.0\%, corresponding to 499 successful translations out of 500 intent instances. This reduced error margin indicates that CAIF provides more reliable validation and better preserves semantic consistency than the general LLM-based baseline.

\begin{table}[h!]
    \centering
    \caption{CAIF LLM Agent Inference Parameters}
    \label{tab:hyperparameters_CAIF}
    \begin{tabular}{ cc }
        \toprule % This will be a bold/thick line at the top
        \textbf{Parameter} & \textbf{Value} \\
        \midrule % This will be a standard mid-rule
        Profiling Agent Model & Qwen3-4B-Instruct-2507 \\
        Evaluator Agent Model & llama-3.3-nemotron-super-49b-v1.5 \\
        Temperature & 0.6 \\
        Top P & 0.95 \\
        \bottomrule % This will be a bold/thick line at the bottom
    \end{tabular}
\end{table}

\begin{figure}
    \centering
    \includegraphics[width=1\linewidth]{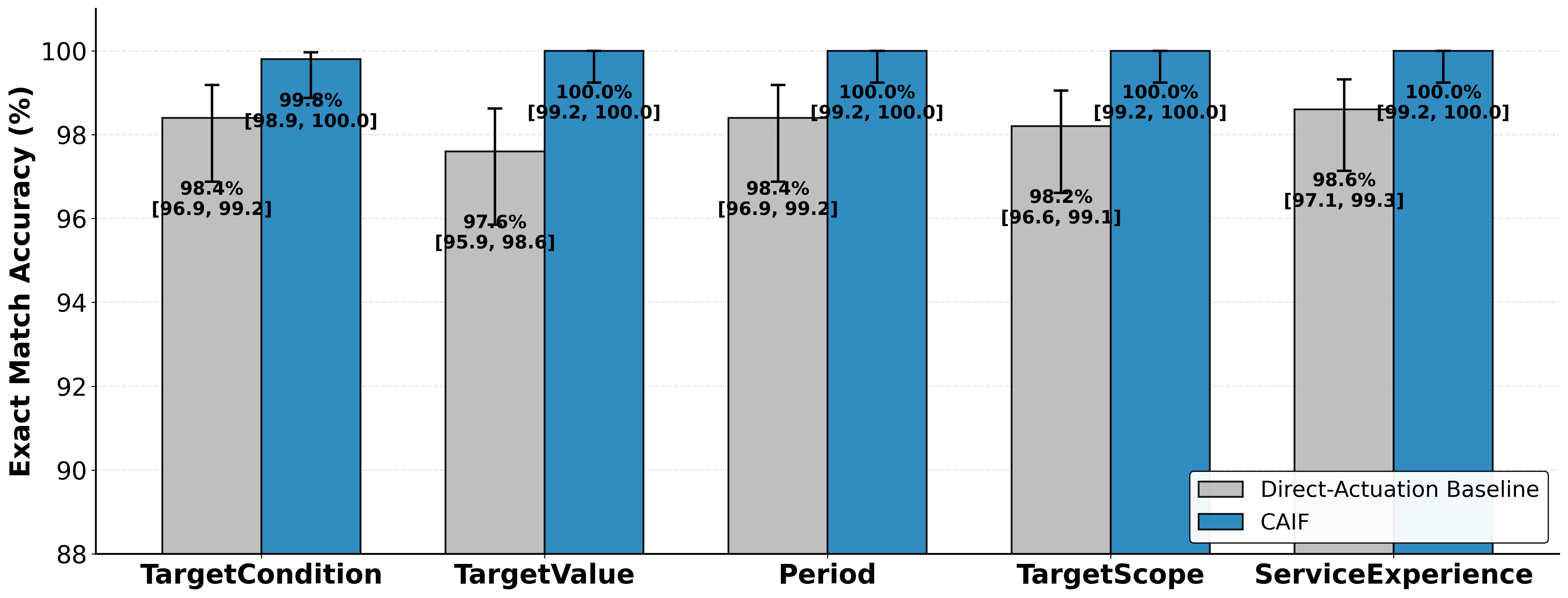}
    \caption{Field-Level Accuracy with 95\% Confidence Interval}
    \label{fig:field_acc}
\end{figure}

\begin{figure}
    \centering
    \includegraphics[width=1\linewidth]{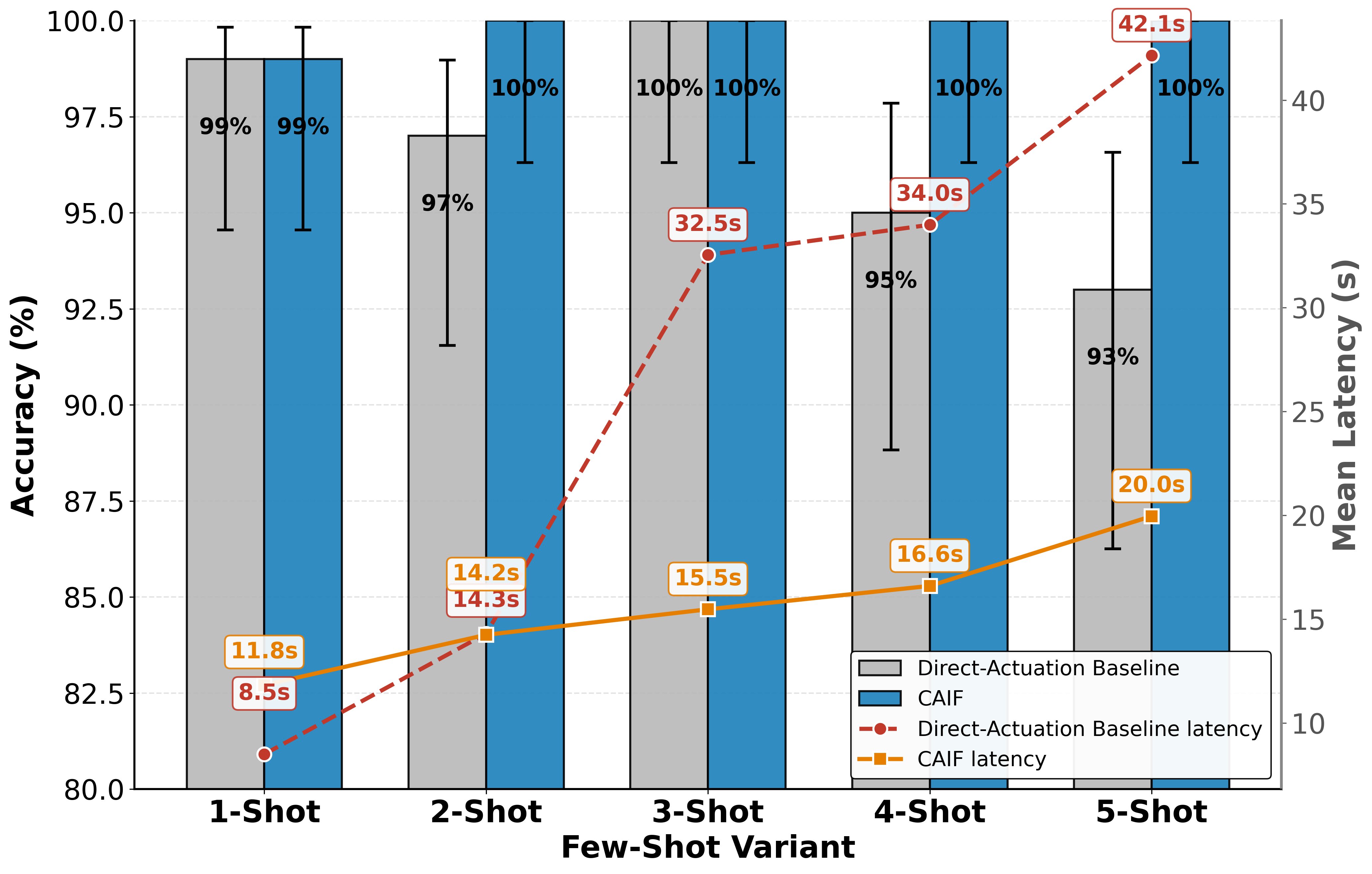}
    \caption{Per-Shot Mean Latency \& Accuracy with 95\% Confidence Intervals }
    \label{shot_latency}
\end{figure}

We conduct a field-level accuracy analysis, as illustrated in Fig. \ref{fig:field_acc}. The results show that CAIF achieves 100\% accuracy with 95\% confidence interval ranging from 99.2\% to 100\% across several key intent attributes, which is much better than direct-actuation baseline. Fig. \ref{shot_latency} presents a per-shot analysis showing that the baseline performance degrades as more iteration required in order for system to fulfill KPIs from intent, particularly in the 4-shot and 5-shot scenarios. We tested the latency, due to the dual-agent coordination CAIF introduces a small overhead in the 1-shot scenario (11.8s vs. 8.5s), but the latency break-even occurs at the 2-shot setting, and CAIF consistently outperforms the baseline in the 3, 4 and 5-shot scenarios.

% \textcolor{blue}{We prepare a set of 200 natural language intent instances. These instances vary in sentence structure, wording, and parameter values (e.g., slice identifiers) to evaluate hallucination and randomness in both architectures. For each intent instance, correctness is defined as whether the generated output semantically matches the original intent and can be successfully executed by the target rApps.}

% \textcolor{blue}{As shown in Figure \ref{fig:SSSO_vs_aLLM}, the direct-actuation baseline architectures achieves an accuracy of 94.5\%. Although most intents are processed correctly, 11 error cases are observed. Among them, one semantic error leads to the generation of an SLA Slice A1 policy that drifts from the original intent, while the remaining ten errors are caused by invalid output formats. Thess incorrect outputs are directly forwarded to the rApps, resulting in execution failures. In contrast, the proposed contract-based framework achieves 100\% accuracy across all 200 intent instances, the contract-based agentic framework successfully detects and corrects the semantic violations or invalid parameters before execution.}
% \begin{figure}
%     \centering
%     \includegraphics[width=1\linewidth]{images/SSSO_vs_aLLM_v2.png}
%     \caption{CAIF Validation}
%     \label{fig:SSSO_vs_aLLM}
% \end{figure}

% \begin{figure}
%     \centering
%     \includegraphics[width=1\linewidth]{images/SSSO_vs_aLLM_v1.png}
%     \caption{Contract-Based Framework Validation}
%     \label{fig:SSSO_vs_aLLM}
% \end{figure}

% \begin{figure}
%     \centering
%     \includegraphics[width=1\linewidth]{images/Token_v1.png}
%     \caption{Contract-Based Framework Token Cost}
%     \label{fig:Token}
% \end{figure}

\subsection{End-to-End Intent-Based Network in SLA Slice Use Case}
% \textcolor{blue}{This section presents the results of our experimental analysis to validate the effectiveness of the proposed contract-based agentic framework approach with network. The experimental environment was established using the VIAVI AI RSG RAN Simulator. Campus environment is considered which was configured to emulate three cellular network with two slices, eMBB (Enhanced Mobile Broadband) and mMTC (Massive Machine Type Communications). In eMBB slice, 10 random walking people with calling (voice) and 30 fixed people for watching the streaming videos are configured. On the other hand, 80 IoT devices are configured in the mMTC slice. The remaining detail parameters are listed in Table \ref{tab:airsgcellconfig} and Table \ref{tab:airsgueconfig}.}

To show the practical utility of the framework, we demonstrate the system in a realistic deployment scenario where the agent manages dynamic network slicing within a simulated environment. This test validates the ability of the contract-based logic to translate high-level intent into specific radio resource management actions within a complex campus network. 

We simulate a campus deployment scenario using the RAN Simulator. This scenario requires three cellular networks consisting of two distinct slices: Enhanced Mobile Broadband (eMBB) and Massive Machine Type Communications (mMTC). The eMBB slice includes ten mobile users engaged in voice calls and thirty stationary users streaming video content. In contrast, the mMTC slice is configured with eighty IoT devices. Detailed network and user equipment parameters for this setup are provided in Table \ref{tab:airsgcellconfig} and Table \ref{tab:airsgueconfig}. To validate the End-to-End functionality from intent formulation to actual network response, we conducted two independent experiments. The first, Single Intent Assurance, verifies the translation of user intent into A1 policies and confirms that the Near-RT RIC closed loop mechanism is correctly triggered. The second, Dynamic Intent Assurance, evaluates the system stability when handling multiple intents arriving at different time intervals.
\begin{table}[h!]
    \centering
    \caption{Scenario Simulation}
    \label{tab:airsgcellconfig}
    \begin{tabular}{ cc }
        \toprule % This will be a bold/thick line at the top
        \textbf{Parameter} & \textbf{Value} \\
        \midrule % This will be a standard mid-rule
        Simulator & VIAVI AI RSG \\
        Version & 2.6 \\
        Number of gNB & 3 \\
        Initial number of UE(s) at gNB$_1$ & 43 \\
        Initial number of UE(s) at gNB$_2$ & 31 \\
        Initial number of UE(s) at gNB$_3$ & 46 \\
        \bottomrule % This will be a bold/thick line at the bottom
    \end{tabular}
\end{table}

\begin{table*}[htbp]
    \centering
    \caption{UE and slice Simulation}
    \label{tab:airsgueconfig}
    \begin{tabular}{ ccccccc }
        \toprule % This will be a bold/thick line at the top
        \textbf{UE Group} & \textbf{Mobility Model} & \textbf{Quantity} & \textbf{Slice} & \textbf{Qos Identifier} & \textbf{Target Throughput} & \textbf{Description} \\
        \midrule % This will be a standard mid-rule
        Pedestrian & Random walk & 10 & eMBB (Slice 1) & 1 (GBR) & 0.5 Mbps &
        Conversation Voice \\
        Person & Fixed & 30 & eMBB (Slice 1) & 8 (Non-GBR) & 40 Mbps &
        Video \\
        IoT Device & Fixed & 80 & mMTC (Slice 2) & 9 (Non-GBR) & 0.25 Mbps &
        IoT Device \\
        \bottomrule % This will be a bold/thick line at the bottom
    \end{tabular}
\end{table*}

\subsubsection{Single Intent Assurance Evaluation}
\begin{figure}
    \centering
    \includegraphics[width=1 \linewidth]{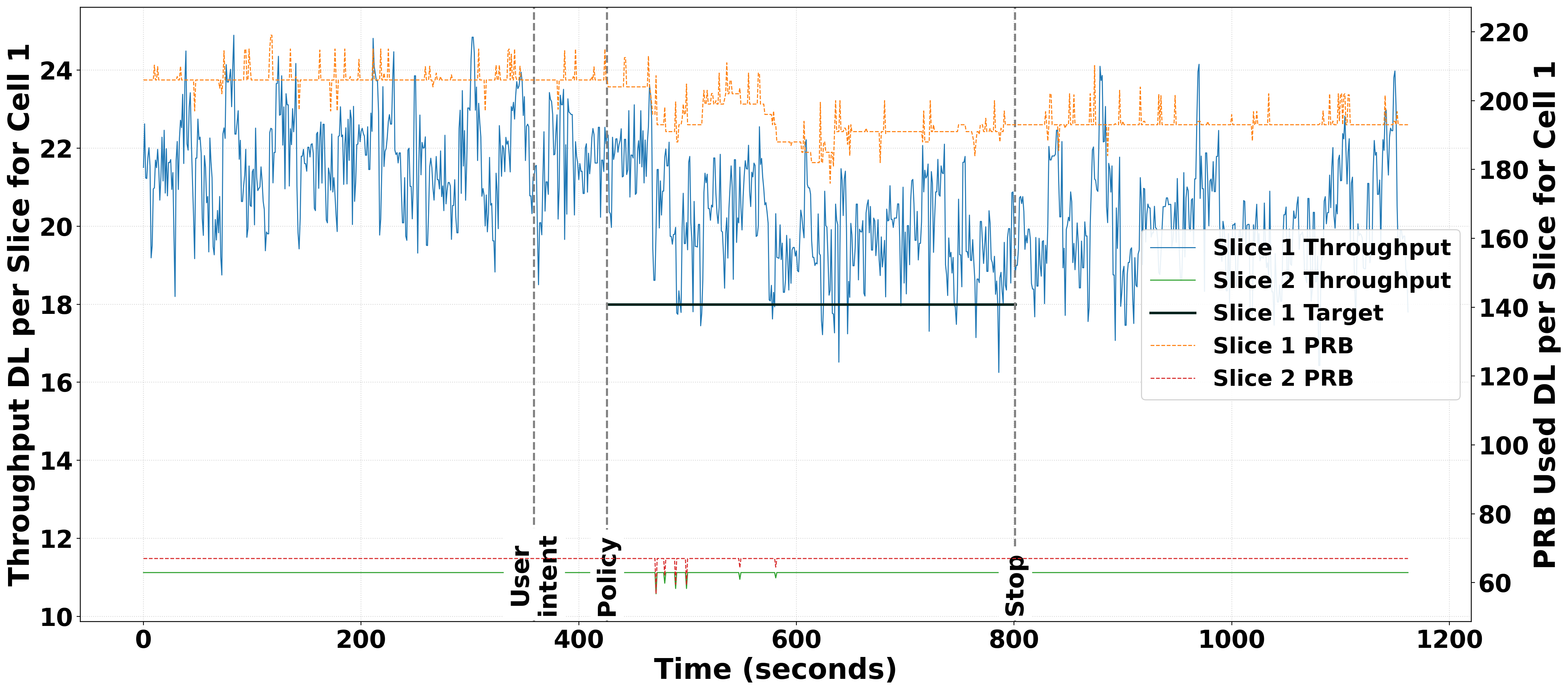}
    \caption{Time series result for single intent assurance in slice 1 of cell 1}
    \label{fig:single_sla_assurance}
\end{figure}

The single intent assurance results for slice 1 are illustrated in Fig. \ref{fig:single_sla_assurance}. When the operator submits a request to "decrease downlink throughput by 20\% in 5 minutes" for slice 1, the system first determines that the current throughput is approximately 22 Mbps. It then calculates the 20\% reduction to set a target of 18 Mbps and prepares the data for the A1 policy.

When Near-RT RIC receives the A1 Policy, xApps start adjusting the rRMPolicyMaxRatio and rRMPolicyMinRatio via E2 interface to maintain the throughput within the 18Mbps every second. After stopping the policy, rRMPolicyMaxRatio and rRMPolicyMinRatio values persist at the final control state and throughput starts to aggressively change.

\subsubsection{Dynamic Intent Assurance Evaluation}
\begin{figure}
    \centering
    \includegraphics[width=1 \linewidth]{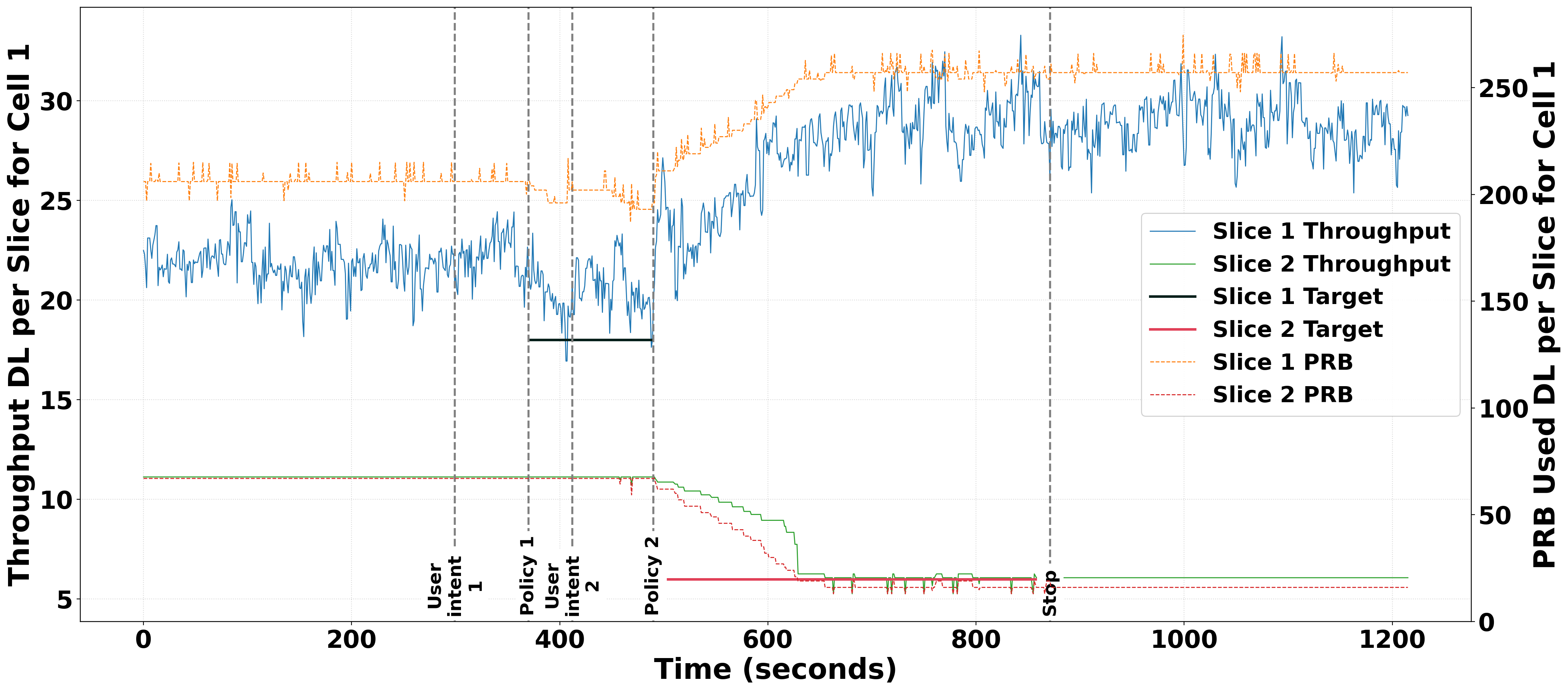}
    \caption{Time series result for dynamic intent assurance from slice 1 to slice 2 of cell 1}
    \label{fig:dynamic_intent_assurance}
\end{figure}

The second test, shown in Fig. \ref{fig:dynamic_intent_assurance} illustrates how the system handles multiple requests from an operator to manage resources across slice 1 and slice 2. When the operator submits the first intent to "decrease downlink throughput by 20\% in 10 minutes" for slice 1, the system identifies that the current throughput is approximately 22 Mbps. It then translates that 20\% reduction into a target of 18 Mbps and prepares the data for the A1 policy.

When the Near-RT RIC receives the A1 policy (indicated by the vertical dashed line labeled “Policy 1”), xApps begin adjusting the rRMPolicyMaxRatio and rRMPolicyMinRatio via the E2 interface to maintain the throughput at 18 Mbps every second. Two minutes later, the operator submits another request: "In slice ID 2 of cell 1, decrease downlink throughput by 50\% in 5 minutes." The system identifies that the current throughput for slice 2 is approximately 12 Mbps, calculates the 50\% reduction to a target of 6 Mbps, and prepares the data for the A1 policy.

When the Near-RT RIC receives this second A1 policy (indicated by the vertical dashed line labeled “Policy 2”), the SLA Slice xApp replaces the first policy. It then maintains slice 2 at 6 Mbps by dynamically adjusting the rRMPolicyMaxRatio and rRMPolicyMinRatio. After the policy stops (indicated by the vertical dashed line labeled “Stop”), the values for these ratios remain at their final controlled levels.

\section{Conclusions and Future Works} \label{Lesson Learned}

This study introduces CAIF, enforced across rApp and xApp workflows for slice resource allocation on an O-RAN testbed. From the experimental evaluation, we identified several critical insights that validate the architecture and motivate directions for future research. First, the results we obtained from our testing show contract-based approach combined with Generative AI for intent translation, significantly outperforms traditional baselines. It effectively acts as a guardrail to prevent minor semantic drifts or format errors can lead to policy execution failures. Future work will expand the evaluation dataset to explore more robust mechanisms for constraining LLM behavior. Second, the End-to-End SLA slice experiments demonstrate that the proposed framework operates effectively in simulated RAN environments with specific scenarios that reflect real-world situations. Both single and dynamic intent assurance scenarios confirm that the framework can continuously adapt network control policies while maintaining SLA requirements. Motivated by these results, future work will extend the proposed framework to support additional RAN management use cases, such as Energy Saving and Traffic Steering.

% \textcolor{blue}{First, the results indicate that relying on a single LLM to directly translate natural language intents into executable network policies is inherently unreliable. Although the baseline architecture using a standalone LLM achieves relatively high accuracy, even minor semantic drifts or format errors can lead to policy execution failures. This highlights that LLM hallucinations and output randomness remain critical risks in direct LLM-based control. Accordingly, future work will explore more robust mechanisms for constraining LLM behavior.}

\section*{Acknowledgment}
This work was supported by the National Science and
Technology Council, Taiwan, under Contract numbers 114-2221-E-011-078-MY3 and 114-2218-E-011-003.

\bibliographystyle{IEEEtran}
\bibliography{bib}

@INPROCEEDINGS{10464635,
  author={Fransiscus Asisi Bimo and Cheng, Ray-Guang and Tseng, Chien-Chao and Chiang, Cheng-Rong and Huang, Chih-Hsiang and Lin, Xiu-Wei},
  booktitle={2023 IEEE Globecom Workshops (GC Wkshps)}, 
  title={Design and Implementation of Next-Generation Research Platforms}, 
  year={2023},
  volume={},
  number={},
  pages={1777-1782},
  doi={10.1109/GCWkshps58843.2023.10464635}}

@misc{bimo2025intentbasednetworkranmanagement,
      title={Intent-Based Network for RAN Management with Large Language Models}, 
      author={Fransiscus Asisi Bimo and Maria Amparo Canaveras Galdon and Chun-Kai Lai and Ray-Guang Cheng and Edwin K. P. Chong},
      year={2025},
      eprint={2507.14230},
      archivePrefix={arXiv},
      primaryClass={cs.NI},
      url={https://arxiv.org/abs/2507.14230}, 
}

@misc{multiagentreference,
      title={Multi-Agent Reinforcement Learning in Wireless Distributed Networks for 6G}, 
      author={Jiayi Zhang and Ziheng Liu and Yiyang Zhu and Enyu Shi and Bokai Xu and Chau Yuen and Dusit Niyato and Mérouane Debbah and Shi Jin and Bo Ai and Xuemin and Shen},
      year={2025},
      eprint={2502.05812},
      archivePrefix={arXiv},
      primaryClass={cs.IT},
      url={https://arxiv.org/abs/2502.05812}, 
}

@ARTICLE{11152698,
  author={Xiao, Yong and Shi, Guangming and Zhang, Ping},
  journal={IEEE Communications Magazine}, 
  title={Toward Agentic AI Networking in 6G: A Generative Foundation Model-as-Agent Approach}, 
  year={2025},
  volume={63},
  number={9},
  pages={68-74},
  doi={10.1109/MCOM.001.2500005}}

@ARTICLE{9814869,
  author={Polese, Michele and Bonati, Leonardo and D'Oro, Salvatore and Basagni, Stefano and Melodia, Tommaso},
  journal={IEEE Transactions on Mobile Computing}, 
  title={ColO-RAN: Developing Machine Learning-Based xApps for Open RAN Closed-Loop Control on Programmable Experimental Platforms}, 
  year={2023},
  volume={22},
  number={10},
  pages={5787-5800},
  doi={10.1109/TMC.2022.3188013}}

@ARTICLE{10574890,
  author={Mekrache, Abdelkader and Ksentini, Adlen and Verikoukis, Christos},
  journal={IEEE Network}, 
  title={Intent-Based Management of Next-Generation Networks: an LLM-Centric Approach}, 
  year={2024},
  volume={38},
  number={5},
  pages={29-36},
  doi={10.1109/MNET.2024.3420120}}

@INPROCEEDINGS{10327837,
  author={Dzeparoska, Kristina and Lin, Jieyu and Tizghadam, Ali and Leon-Garcia, Alberto},
  booktitle={2023 19th International Conference on Network and Service Management (CNSM)}, 
  title={LLM-Based Policy Generation for Intent-Based Management of Applications}, 
  year={2023},
  volume={},
  number={},
  pages={1-7},
  doi={10.23919/CNSM59352.2023.10327837}}

@ARTICLE{9443201,
  author={Abbas, Khizar and Khan, Talha Ahmed and Afaq, Muhammad and Song, Wang-Cheol},
  journal={IEEE Access}, 
  title={Network Slice Lifecycle Management for 5G Mobile Networks: An Intent-Based Networking Approach}, 
  year={2021},
  volume={9},
  number={},
  pages={80128-80146},
  doi={10.1109/ACCESS.2021.3084834}}

@INPROCEEDINGS{10539172,
  author={Manias, Dimitrios Michael and Chouman, Ali and Shami, Abdallah},
  booktitle={2024 20th International Conference on the Design of Reliable Communication Networks (DRCN)}, 
  title={Towards Intent-Based Network Management: Large Language Models for Intent Extraction in 5G Core Networks}, 
  year={2024},
  volume={},
  number={},
  pages={1-6},
  doi={10.1109/DRCN60692.2024.10539172}}

@misc{tmf,
title ={TMF921 Intent Management API User Guide},
author= {TMFORUM},
url ={https://www.tmforum.org/resources/specifications/tmf921-intent-management-api-user-guide-v5-0-0/},
lastaccessed = "June 15, 2025",
}

@INPROCEEDINGS{10298342,
  author={Habib, Md Arafat and Zhou, Hao and Iturria-Rivera, Pedro Enrique and Elsayed, Medhat and Bavand, Majid and Gaigalas, Raimundas and Ozcan, Yigit and Erol-Kantarci, Melike},
  booktitle={2023 IEEE 20th International Conference on Mobile Ad Hoc and Smart Systems (MASS)}, 
  title={Intent-driven Intelligent Control and Orchestration in O-RAN Via Hierarchical Reinforcement Learning}, 
  year={2023},
  volume={},
  number={},
  pages={55-61},
  doi={10.1109/MASS58611.2023.00015}}

@misc{intent_dataset,
  author = {Fransiscus Asisi Bimo and Chun-Kai Lai and Zhi-Yuan Yang and Ray-Guang Cheng},
  title = {{**Contract-
based Agentic Intent Framework**}},
  year = {2026},
  howpublished = {GitHub},
  url = {https://github.com/bmw-ece-ntust/Contract-based_Agentic_Intent_Framework},
  note = {Accessed: Feb. 28, 2026}
}

\vspace{12pt}

\end{document}